\documentclass[twocolumn,prb,showpacs,amsmath,amssymb,superscriptaddress,floatfix]{revtex4} 
\usepackage{graphicx}
\usepackage{dcolumn}% Align table columns on decimal point
\usepackage{bm}% bold math
\newcommand{\beq}{\begin{equation}}
\newcommand{\eeq}{\end{equation}}
\newcommand{\beqa}{\begin{eqnarray}}
\newcommand{\eeqa}{\end{eqnarray}}
\newcommand{\kvec}{{\bf k}}
\newcommand{\qvec}{{\bf q}}
\newcommand{\QVEC}{{\bf Q}}
\newcommand{\rvec}{{\bf r}}

\begin{document}

\title{Charge instabilities and electron-phonon interaction in the
  Hubbard-Holstein model}
 \author{A. Di Ciolo}
 \affiliation{Dipartimento di Fisica, Universit\`a di Roma
 ``La Sapienza'', P.  Aldo Moro 2, 00185 Roma, Italy.}
 \affiliation{SMC-Istituto Nazionale di Fisica della Materia}
 \author{J. Lorenzana}
 \affiliation{Dipartimento di Fisica, Universit\`a di Roma
 ``La Sapienza'', P. Aldo Moro 2, 00185 Roma, Italy.}
 \affiliation{SMC-Istituto Nazionale di Fisica della Materia}
 \affiliation{ISC-Consiglio Nazionale delle Ricerche} 
 \author{M. Grilli}
 \affiliation{Dipartimento di Fisica, Universit\`a di Roma
 ``La Sapienza'', P.  Aldo Moro 2, 00185 Roma, Italy.}
 \affiliation{SMC-Istituto Nazionale di Fisica della Materia}
 \author{G. Seibold}
 \affiliation{Institut f\"ur Physik, BTU Cottbus, PBox 101344,
          03013 Cottbus, Germany.}

\date{\today}% It is always \today,
\begin{abstract}
We consider the Hubbard-Holstein  model in the adiabatic limit
to investigate the effects of electron-electron interactions on the electron-phonon
coupling. To this aim we compute at any momentum and filling
 the static charge susceptibility of the 
Hubbard model within the Gutzwiller approximation and we find that
electron-electron correlations effectively screen the electron coupling to the
lattice. This screening is more effective at large momenta and, as a consequence,
the charge-density wave phase due to the usual Peierls instability of the
Fermi surface momenta is replaced by a phase-separation instability 
when the correlations are sizable.
\end{abstract}
\maketitle
\section{Introduction}
\label{sec:int}
In the last  years the issue of electron-phonon ($e$-$ph$) coupling in the presence
of strong electron-electron ($e$-$e$) correlations has been raised in a variety 
of contexts. For instance, in the
high-temperature superconducting cuprates recent photoemission
experiments\cite{Lan01, DAs02, Gwe04} 
indicate a sizable coupling of electrons with collective modes, possibly 
of phononic nature, while the softening of a phonon peak 
in inelastic neutron scattering experiments,~\cite{Rez06,Rez07} as well as features
in tunnelling spectra\cite{lee06} suggest that electrons are substantially coupled
to the lattice in these materials. At the same time optical and transport experiments do not
display a strong $e$-$ph$ coupling except at very small doping, where polaronic features
have been observed.~\cite{polaronexper1,polaronexper2,polaronexper3} 
An intriguing interplay between lattice and electrons
has also been invoked to explain transport in manganites\cite{Mil96}, in single-molecule
junctions\cite{Grempel} and in fullerenes,~\cite{Gun97} where a correlation-enhanced
superconductivity has also been proposed.~\cite{capone}
These examples show that the issue of $e$-$ph$ coupling in the presence of strong
$e$-$e$ correlation is generally relevant and it translates in several related issues.
First of all, the fact that various physical quantities appear to be differently
affected by phonons indicates that the energy and momentum structure of the
$e$-$ph$ coupling is important. In turn this emphasizes the
role of  $e$-$e$ interactions as an effective mechanism 
to induce strong energy and momentum dependencies in the $e$-$ph$ coupling.
Secondly, phonons may be responsible for charge instabilities. One possibility 
is that they mediate interactions between electrons on the Fermi surface giving rise
to charge-density waves or Peierls distortions. It has also been proposed\cite{Gri94,CDCG95}
that a phonon-induced attraction gives rise to an electronic phase separation (although  in 
real systems this is ultimately prevented by the long-ranged Coulombic forces with the formation
 of nano- or mesoscopic domains, the so-called frustrated phase
 separation.~\cite{low94,lor01I,lor01II,lor02,Ort06,Ort07,ort08}) 

Due to the above, phonons coupled to strongly correlated electrons have already been investigated
by means of  numerical techniques like quantum Monte
Carlo,~\cite{HS83, hir83, HS84, Hir85, Ber95, Hua03} 
exact diagonalization,~\cite{dob94,dob94epl,lor94b,stephan}  Dynamical
Mean Field Theory (DMFT),~ \cite{FJ95, Capo04, Kol04-1, Kol04-2, Jeo04,
  San05, San06} and  (semi)analytical approaches like slave bosons (SB) and large-N expansions.~
\cite{Gri94, Kel95, Zey96, Koc04, Cap04, Cit01, Cit05} Despite this variety of approaches,
a systematic and thorough investigation within the same technical framework is not yet available
either due to the demanding character of the numerical approaches or to the limited
parameter ranges investigated so far. Therefore
in this paper we study the renormalization of the electron-lattice coupling  in the presence of 
strong $e$-$e$ correlations  systematically considering the momentum, doping, and interaction-strength
dependencies. In particular we want to elucidate how charge density wave (CDW) or  
phase separation (PS) instabilities are modified in the presence of $e$-$e$ interactions. 
To this aim we need a technique which is not numerically very demanding, but
still provides a quantitatively acceptable treatment of the strongly correlated regime. In this regard
we find the Gutzwiller approach and the related Gutzwiller Approximation\cite{gut63} (GA) a good compromise
allowing extensive and systematic exploration of various parameter ranges while keeping a reliable
treatment of the low-energy physics. It has recently been shown that
the Gutzwiller variational approach provides remarkably accurate positions of 
complex magnetic phase boundaries in infinite
dimensions.~\cite{gun07} This indicates that the Gutzwiller energy and
its derivatives are quite accurate. In this work we extend these results to
the charge channel also in infinite dimensions, where the GA to
the Gutzwiller variational problem is exact.~\cite{met88,geb90} In addition, in order to
make contact with layered systems, we study the two-dimensional (2d)
case where the GA is still expected to give an accurate estimate of
the energy. This also gives us the opportunity to study the interplay
between nesting in the presence of electron phonon coupling, 
which favors Peierls distortions, and strong
correlation which favors phase separation. 

To obtain the phase diagram in the
presence of both  $e$-$e$ and $e$-$ph$ interactions,  in principle, one should compute the GA
energy for every possible charge-ordered state. 
However, it is much
more practical to study the static response functions of the uniform
state to an external perturbation and to locate the relevant instabilities. 
We show below that for
Holstein phonons in the adiabatic limit, the {\em exact} charge
susceptibility in the presence of  $e$-$ph$ and  $e$-$e$ interaction
is simply related to the charge susceptibility without
phonons. Therefore our work reduces to compute the latter which is done
in the GA. This corresponds to the static
limit of the GA+(random phase approximation), (GA+RPA), which was derived in
Ref.~\onlinecite{sei01} and is rooted in  Vollhardt's Fermi liquid
approach.~\cite{vol84} As a byproduct our work generalizes Vollhardt's
computation of the zero-momentum and half-filled charge susceptibility
to  any momentum and filling. 
 Our approach is not as accurate as QMC or DMFT studies as far as the
electron dynamical excitations are concerned, because it inherently deals with the quasiparticle
part of their spectrum. Nonetheless with relatively small numerical efforts it allows for
a systematic analysis of momentum, doping and interaction dependencies of the screening
processes underlying the quasiparticle charge response and the related $e$-$ph$
coupling. 

It is worth mentioning that the dynamical version of our GA+RPA approach
has been tested in various situations and found to be accurate
compared with exact diagonalization.~\cite{sei01,sei03,sei04b,sei07b} 
Computations for realistic models have provided a description of
different physical quantities in accord with
experiment.~\cite{lor02b,lor03,sei05}  The present analysis of the
charge susceptibility of the paramagnetic state gives us an opportunity
to present the  method in a simpler context with respect to the 
more complex situations considered in the past, thus allowing for the
clarification of several methodological aspects. 
On the other hand, restricting to the paramagnetic state we ignore
antiferromagnetic and related instabilities that arise as the system approaches
half-filling.

The scheme of our paper is as follows.  
We first present in Sect. II the derivation of an exact result 
for the renormalization of the $e$-$ph$ 
coupling in the adiabatic limit. The GA+RPA approach is presented in Sect. III and then it is 
exploited to systematically calculate the momentum, doping, and interaction
dependencies of the charge susceptibility of the Hubbard model. The results in $d=\infty$ are contained
in Sect. IV, while the 2d case is reported in Sect. V.
Our conclusions can be found in Sect. VI, while the details of our calculations are given in the Appendix. 

\section{Exact relation between electron-phonon instabilities and charge susceptibility}
%\label{sec:for}
In this section we show the renormalization of the $e$-$ph$ coupling and how the renormalized coupling is 
related to the electronic susceptibility. 
We consider a single band system with $e$-$e$ interaction and $e$-$ph$ coupling on a lattice
\begin{eqnarray}
H_{tot}=H_{e}+H_{eph}+H_{ph}
\label{Htot}
\end{eqnarray}
Specifically we consider the  Holstein interaction%\cite{Hol59}
\begin{eqnarray}
H_{eph}+H_{ph}= \sum_{i}\beta x_{i}(\hat{n}_{i}-n) + 
\sum_{i}\left(\frac{P_{i}^2}{2M} + \frac{1}{2}K x_{i}^2\right)
\label{Hh1}
\end{eqnarray} 
where $\hat{n}_i$ is the electronic density operator; $n=N_{p}/N$ is the 
average density on a lattice with $N$ sites and $N_p$ particles.  
$P_i$, $x_i$ and $M$ are the momentum, the displacement and the mass of the lattice ions
respectively.
We treat Eq.~(\ref{Hh1}) in the extreme adiabatic limit ($M$$\rightarrow$$\infty$), where, using the 
Born-Oppenheimer's principle, we can represent the ground state of the system as 
$|\psi_{tot}\rangle = |\psi_{e}\rangle|\chi_{ph}\rangle$. 
The electronic wave function $|\psi_{e}\rangle$ depends parametrically 
on the ionic displacement $x_i$. We assume that the ground state in
the absence of $e$-$ph$ coupling is uniform [{\it i.e.}, no static 
CDW state is present]. For a fixed configuration of displacements 
$\{x_i\}$, the $e$-$ph$ term of Eq.~(\ref{Hh1}) acts as an external field 
on the electrons, producing a density deviation $\{\delta n_i\}$,
being $\delta n_{i}=\langle \hat{n}_{i} \rangle - n$. 

The total energy in the adiabatic limit has the form 
\begin{eqnarray}
E_{tot} = E_{e}[\delta n] + E_{eph}[\delta n,x] + E_{ph}[x]
\label{Etot}
\end{eqnarray} where $E_{e}=\langle \psi_{tot}|H_{e}|\psi_{tot}\rangle$, $E_{eph}=\langle \psi_{tot}|H_{eph}|\psi_{tot}\rangle$ and $E_{ph}=\langle \psi_{tot}|H_{ph}|\psi_{tot}\rangle$, and $\delta n$, $x$ stand for the sets $\{\delta n_i\}$ and $\{x_i\}$ respectively. 
We move to momentum space and perform an expansion of 
$E_{e}[\delta n]$ up to the second order in the density deviation 
$\delta n$ 
\begin{widetext}
\begin{eqnarray}
E_{e}[\delta n] = E_{e}^{(0)} +  \sum_{\qvec} \left(
\frac{\partial E_{e}}{\partial n_{\qvec}}\right)_0 \delta n_{\qvec} +  
\frac{1}{2}\sum_{\qvec} \left(\frac{\partial^2 E_{e}}{\partial n_{\qvec}\partial n_{-\qvec}}
\right)_0\delta n_{\qvec}\delta n_{-\qvec},
\label{Eeq}
\end{eqnarray}
\end{widetext}
where the label ``$0$'' indicates that the derivatives must be evaluated
within the uniform electronic state (in the absence of the
$e$-$ph$ coupling); $E_{e}^{(0)}$ is therefore the ground state energy in the absence of the $e$-$ph$ coupling. The first order term vanishes
identically. For the $\qvec \neq 0$ terms this arises because 
stability requires that odd powers in the $\delta n_\qvec$  expansion must vanish (otherwise the systems would lower its energy by creating a CDW state).
Instead the $\qvec=0$ term with $\partial E_{e} / \partial n_{\qvec=0}$ vanishes because we are working at
a fixed particle number ($\delta n_{\qvec=0}=0$). 
Therefore the electronic ground state energy is quadratic in the density deviation 
\begin{equation}
E_{e}[\delta n_{i}]= E_{e}^{(0)} + \frac{1}{2} \sum_{\qvec}\kappa^{-1}_{\qvec}
\delta n_{\qvec}\delta n_{-\qvec}
\label{Eeq1}
\end{equation}
where we customarily define 
$\kappa_{\qvec}^{-1}\equiv 
\left(\frac{\partial^2E_{e}}{\partial \delta n_{\qvec}^2}\right)_0$
with  $\kappa_{\qvec}$ being the static charge susceptibility of the electronic system in the absence of the $e$-$ph$ coupling. 

We now minimize the Holstein energy
\begin{eqnarray}
 E_{H}=E_{eph}[\delta n,x] + E_{ph}[x] = 
\sum_{i}\left(\beta x_{i}\delta n_{i} + \frac{1}{2}K x_{i}^2\right)
\label{Ett}
\end{eqnarray} 
with respect to the ionic displacements $x_i$ at fixed $\delta n_i$
finding $x_{i}$=$-\beta \delta n_{i}/K$. Replacing this expression 
in Eq.~(\ref{Ett}) and introducing the adimensional coupling 
\begin{equation}
  \label{eq:lambda}
\lambda \equiv \frac{\chi^{0}_0\beta^2}K
\end{equation}
[$\chi^{0}_0$ is the density of states (DOS) of the non interacting electron system], we find the following expression for $E_{tot}$  

\begin{eqnarray}
E_{tot} = E^{(0)}_{tot} + \frac{1}{2}\sum_{\qvec}\left(\kappa^{eph}_{\qvec}\right)^{-1}\delta n_{\qvec}\delta n_{-\qvec}
\label{E2}
\end{eqnarray} with
\begin{eqnarray}
\kappa^{eph}_{\qvec}=\frac{\kappa_\qvec}{1-\lambda\kappa_\qvec/\chi^0_0}=\frac{\kappa_\qvec}{1-\tilde{\lambda}_\qvec}
\label{kep}
\end{eqnarray}
where we introduced the renormalized coupling
\begin{equation}
  \label{eq:lambdaren}
 \tilde{\lambda}_\qvec \equiv \frac{\lambda \kappa_\qvec}{\chi^0_0}. 
\end{equation}

Eqs.~(\ref{E2})-(\ref{eq:lambdaren}) provide the exact second-order expansion of the total energy in
the adiabatic limit and establish a relation between the electronic
charge susceptibilities, $\kappa_{\qvec}$ and $\kappa^{eph}_{\qvec}$,
in the absence and in the presence of the $e$-$ph$ coupling
respectively. By construction $\tilde{\lambda}_\qvec$ is defined in
such a way that $\tilde{\lambda}_\qvec=1$ indicates an instability, in
analogy with the noninteracting $e$-$ph$ coupling for which
$\lambda=1$ indicates a $\qvec=0$ instability.

We can write $\tilde{\lambda}_\qvec = \beta \Gamma_\qvec
\kappa^0_\qvec / K$, using the 
charge vertex\cite{Kel95, Hua03, Koc04, Cap04, Cit01, Cit05}
which acts as a renormalized {\em quasiparticle}-$ph$ coupling 
\begin{equation}
  \label{eq:gamma}
  \Gamma_\qvec \equiv \frac{ \kappa_\qvec}{\kappa^0_\qvec}\beta
\end{equation}
 ($\kappa^0_\qvec$ is the non interacting susceptibility). Then we find 
\begin{eqnarray}
\frac{\tilde{\lambda}_\qvec}{\lambda} = \frac{\kappa^0_\qvec}{\chi^0_0}\frac{\Gamma_\qvec}{\beta}
\label{lbdq}
\end{eqnarray}
Eq.~(\ref{lbdq}) has been introduced to separate in the $e$-$ph$ coupling
the effects of finite $\qvec$ from those of the $e$-$e$ renormalization.
Specifically, $\kappa^0_\qvec / \chi^0_0$ contains the effects
of finite momentum, and is present even for non-interacting electrons,
while the $e$-$e$ interaction acts on the $e$-$ph$
coupling via a modification of the electronic charge susceptibility, given by $\Gamma_\qvec / \beta$.
In this exact adiabatic derivation both these effects act
in a simple multiplicative manner on $\lambda$.

A second feature of the result in Eq.~(\ref{kep}) is that
the system can become unstable if, upon increasing $\lambda$, it meets
the condition $\lambda=\lambda_c$ with 
\begin{equation}
\label{eq:lambdac}
\lambda=\lambda_c=\chi^0_0/\kappa_{\qvec_c}  
\end{equation}
for some $\qvec_c$. In this case, if no other (first-order)
instabilities take place before, the system undergoes a transition to
a charge-ordered state with a typical wavevector $\qvec_c$. 

Now the whole issue to study the effects of $e$-$e$ interactions on the
$e$-$ph$ coupling and the related electronic charge instability is
reduced to the study of the electronic charge susceptibility (in the
absence of the $e$-$ph$ interaction). This is the main goal of the subsequent
sections.

%\emph{}

%In particular, we first describe our GA+RPA formalism in Section III, then we discuss our results for different dimensions: in 1d (Section IV), 2d (Section V) and d=$\infty$ (Section VI). Last we present our general conclusions in Section VII and the details of our calculations are given in the Appendix. 

\emph{}

\section{Formalism}
%\label{sec:for}

In order to compute the static charge susceptibility, we evaluate the electronic energy in the presence of an external field $f$ 
\beqa
H_{f} = \sum_{ij\sigma}f_{ij\sigma} c^{\dag}_{i\sigma}c_{j\sigma}
\label{Hf}
\eeqa 
where
$c^{\dag}_{i\sigma}$($c_{i\sigma}$) are creation (annihilation) fermionic operators and $f_{ij}=f^{*}_{ji}$. 

Insofar our treatment of the $e$-$e$ interactions has been general. 
Starting from this section, we will adopt the one-band Hubbard model with hopping $t_{ij}$ extended to whatever neighbors $i$,$j$ 
\begin{eqnarray}
H_e =\sum_{ij\sigma}t_{ij}c^{\dag}_{i\sigma}c_{j\sigma} + \sum_{i}U \hat{n}_{i\uparrow}\hat{n}_{i\downarrow}
\label{H}
\end{eqnarray}
$\hat{n}_{i\sigma}$=$c^{\dag}_{i\sigma}c_{i\sigma}$ is the density operator associated to the operators $c^{\dag}_{i\sigma}$, $c_{i\sigma}$ and $U$ is the on-site Hubbard repulsion. 
In the numerical computations below we consider only nearest neighbor hopping $t_{ij}=-t$ to be nonzero.  

We apply the Gutzwiller (GZW) variational method to Eq.~(\ref{H}) on
d-dimensional hypercubic lattices with lattice parameter $a$=1. We
consider the GZW ansatz state $|\psi\rangle=\hat{P}|Sd\rangle$, where
the GZW projector\cite{geb90} $\hat{P}$ acts on the Slater determinant
$|Sd \rangle$. Although we analyze a paramagnetic uniform 
state in order to determine the stability we need the energy in the
presence of an arbitrary perturbation of the charge thus  
$|Sd \rangle$ shall allow for broken symmetries. 

The GZW variational problem can not be solved exactly except for particular cases, as for $d$=$\infty$.
Thus one uses the GA. In particular, we use the energy functional
$E_{e}[\rho, D]$ obtained by Gebhard\cite{geb90} which is equivalent to the Kotliar-Ruckenstein saddle point energy \cite{Kot86}
\beqa
E_{e}[\rho,D] = \sum_{ij\sigma} t_{ij}z_{i\sigma}z_{j\sigma} \rho_{ji\sigma} + \sum_{i}UD_i
\label{egeb}
\eeqa
with the GA hopping factors $z_{i\sigma}$  
\begin{widetext}
\beq
z_{i\sigma}[\rho, D] = \frac{\sqrt{(1-\rho_{ii}+D_i)(\rho_{ii\sigma}-D_i)}+\sqrt{D_{i}(\rho_{ii,-\sigma}-D_{i})}}{\sqrt{\rho_{ii\sigma}(1-\rho_{ii\sigma})}}
\label{zis}
\eeq
\end{widetext}being $\rho_{ii}$=$\sum_{\sigma}\rho_{ii\sigma}$. 
Here $\rho$ is the single fermion density matrix in the uncorrelated state  $\rho_{ji\sigma \sigma^{'}}=\langle Sd| c^{\dag}_{i\sigma}c_{j\sigma^{'}}|Sd \rangle$. $D$ is the vector of the GA double occupancy parameters $D_{i}=\langle \psi| \hat{n}_{i\uparrow}\hat{n}_{i\downarrow}|\psi\rangle$.  

To consider arbitrary deviations, the charge and spin distribution
 $\rho$ and the set of $D$ should be completely unrestricted.
Since we will consider essentially charge deviations only 
the part diagonal in the spin indexes
 contributes, therefore we will use the notation  $  \rho_{ij\sigma}\equiv\rho_{ij\sigma \sigma}$.

One can show that the expectation value of the diagonal elements $\rho_{ii\sigma}$, calculated for the $|Sd\rangle$, coincides with the value of the density $n_{i\sigma}$, calculated for the GZW state $|\psi\rangle$.\cite{geb90} Namely, 
\begin{eqnarray}
 n_{i\sigma}= \langle \psi|{c}^{\dag}_{i\sigma}c_{i\sigma}|\psi\rangle = \rho_{ii\sigma} = \langle Sd|{c}^{\dag}_{i\sigma}c_{i\sigma}|Sd\rangle  
\label{nr}
\end{eqnarray}
Eq.~(\ref{nr}) will permit us to express charge density deviations $\delta n_{i} = \sum_{\sigma}\delta n_{i\sigma}$ as appearing in Eq.~(\ref{Etot}) via $\delta \rho_{ii} = \sum_{\sigma}\delta \rho_{ii\sigma}$. 

We find the saddle point solution minimizing Eq.~(\ref{egeb}) with respect to $\rho$ and $D$. 
The variation with respect to $\rho$ has to be constrained to the subspace of the Slater determinants by imposing the projection condition $\rho$=$\rho^2$ 
\beq
\delta \left \{E_{e}[\rho, D] - Tr[\Lambda (\rho^{2}-\rho)] \right \}=0
\label{egav}
\eeq
$\Lambda$ is the Lagrange parameter matrix. 
Then it is convenient to define a GZW Hamiltonian ${h}[\rho,D]$ \cite{rin80, bla86}
\beq
{h}_{ij\sigma}[\rho,D]=\frac{\partial E_{e}}{\partial \rho_{ji\sigma}}
\label{hgzw}
\eeq
The variation of Eq.~(\ref{egav}) with respect to $\rho$ leads to
${h} - \rho \Lambda - \Lambda \rho + \Lambda = 0$
The Lagrange parameters can be eliminated algebraically.\cite{bla86} 
Considering also the variation with respect to $D$, we obtain the self-consistent GA equations 
\beqa
[{h},\rho]_{-}=0,      
\label{statmot}
\eeqa
\beqa               
\frac{\partial E_{e}[\rho,D]}{\partial D_{i}}=0.
\label{hgz}
\eeqa
Eq.~(\ref{statmot}) implies that at the saddle point, $h$ and $\rho$
can be simultaneously diagonalized by a  transformation 
of the single fermion orbital basis 
\beq
c_{i\sigma}=\sum_{\nu}\Psi_{i\sigma}^{\nu}a_{\nu}.
\label{lintra}
\eeq
leading to a diagonal ${h_0}$,  ${h}_{0\mu\nu}=\delta_{\mu\nu}\epsilon_{\nu}$. 
Moreover, the diagonalized $\rho_0$ has an eigenvalue 1 for states
below the Fermi level (hole states) and 0 for states above the Fermi
level (particle states).  We use a 0 to distinguish quantities evaluated at the
saddle point. 

In the absence of an external field, we will consider the paramagnetic homogeneous state as the saddle point solution, {\em i.e.} we expand the energy around the paramagnetic saddle point, which we describe here.  
Starting from a system with density $n=1-\delta$ ($\delta$ is the doping)
and introducing the notation by Vollhardt, one finds
for the GA hopping factors \cite{vol84}
\begin{eqnarray}
z_0 &=& \sqrt{\frac{2x^2-x^4-\delta^2}{1-\delta^2}}\label{z0}, \\
x &=& \sqrt{1-n+D}+\sqrt{D}. \label{xd}
\end{eqnarray}
For the GA energy one obtains, 
\begin{eqnarray}\label{EGAK}
E_{e0}&=&N z_0^2 e^0 + N U D, \\
e^0&=&\int_{-\infty}^{\mu}\!\!d\omega \,\, \omega \rho^0(\omega) 
\end{eqnarray}
where $e^0$, $\rho^0$, and $\mu$  denote the energy per site, the
density of states and the Fermi energy of the non-interacting system
respectively.

The minimization of Eq.~(\ref{EGAK}) yields
\begin{equation}
\frac{x^4(1-x^2)}{x^4-\delta^2}=(1-\delta^2)
\frac{U}{8|e^0|} \equiv u,
\end{equation}
which, by using Eq.~(\ref{xd}), determines the double-occupancy parameter $D_0$.
 
Within the subspace of the Slater determinants, we now consider small amplitude deviations of the density matrix $\rho$ due to $H_f$, given in Eq.~(\ref{Hf}).  This leads to an additional contribution $E_{f}[\rho]$ to Eq.~(\ref{egeb}) 
\beq
E_{f}[\rho]=\sum_{ij\sigma}f_{ij\sigma}\rho_{ji\sigma}
\label{ef}
\eeq 
The field $f$ produces small amplitude deviations $\delta \rho$, $\delta D$ around the unperturbed saddle point density, {\em i.e.} $\delta \rho = \rho -\rho_0$ and $\delta D = D-D_0$: $\delta \rho$ and $\delta D$ are both linear in $f$. In the presence of the external field $f$, Eq.~(\ref{statmot}) will turn into 
\beqa
[{h}+f,\rho]=0      
\label{pertmot}
\eeqa
We will expand $E_{e}^{GA}[\rho,D] = E_{e}[\rho,D] + E_{f}[\rho]$  around the saddle point $E_{e0}$ up to the second order in $\delta \rho$ and $\delta D$\cite{sei01,sei03}
\beqa
E_{e}^{GA}[\rho,D] = E_{e0} + \delta E_{e}^{(1)} + \delta E_{e}^{(2)}. 
\label{exp}
\eeqa
$\delta E_{e}^{(1)}$ ($\delta E_{e}^{(2)}$) contains first (second) order derivatives of the GA energy. We will first consider $\delta E_{e}^{(1)}$ in this section and then $\delta E_{e}^{(2)}$ in the two following subsections. 

The expression for $\delta E_{e}^{(1)}$ is
\begin{eqnarray}
\delta E_{e}^{(1)} = Tr[{h}_{0}\delta \rho] + Tr[f\delta \rho]
\label{Trhf}
\end{eqnarray}   
It is convenient to work in the momentum space, where
${h}_{0\kvec\sigma,
  \kvec'\sigma'}=\delta_{\kvec\kvec'}\delta_{\sigma\sigma'}\epsilon_{\kvec\sigma}$. In addition we restrict the external perturbation to a local field on the charge sector: $f_{ij\sigma}=\delta_{ij}f_i$ with $\sum_i f_i=0$ so that
Eq.~\eqref{ef} becomes 
\begin{equation}
  \label{eq:ef}
E_f[\rho]=\sum_{\qvec}f_{-\qvec} \delta\rho_{\qvec}
\end{equation}
where we introduced the Fourier transform of the density deviation
$\delta \rho_\qvec$  
\begin{eqnarray} 
\delta \rho_{\qvec} = \frac{1}{N}\sum_{\kvec\sigma}\delta \rho_{\kvec+\qvec,\kvec\sigma}.
\label{drqk}
\end{eqnarray}

 We will call  unoccupied states as particle states ($p$) and use the short hand notation $k>k_{F}$ for the restriction in the momentum; analogously  hole ($h$) state are occupied with $k < k_{F}$, $\kvec_{F}$ being the Fermi momentum.

The matrix elements of the $\delta \rho$ are not all independent\cite{bla86}
since $\rho$ must fulfill the projector condition $\rho^2=\rho$ which we can write in terms of $\delta \rho$ 
\beq
\delta \rho = \rho_{0}\delta \rho + \delta \rho \rho_{0} + (\delta \rho)^2 
\label{drd}
\eeq
Since $\rho_{0\kvec\sigma\kvec'\sigma'}=\delta_{\kvec\kvec'} \delta_{\sigma\sigma'}\rho_{\kvec}$ with $\rho_\kvec=1$ for
  $k<k_f$ and 0 otherwise $\rho_{0}$  projects onto occupied
  states.  

Taking matrix elements of Eq.~\eqref{drd} one finds for $hh$ density
deviations  ($k,k' < k_F$) 
\beq
\delta \rho_{\kvec\sigma\kvec'\sigma'}=-\sum_{k'' > k_F,\sigma''}\delta \rho_{\kvec\sigma\kvec''\sigma''}\delta \rho_{\kvec''\sigma''\kvec'\sigma'}\label{drhh}
\eeq
and for the $pp$ density deviations ($k,k' > k_F$) 
\beq
\delta \rho_{\kvec\sigma\kvec'\sigma'}=\sum_{k'' < k_F,\sigma''}\delta \rho_{\kvec\sigma\kvec''\sigma''}\delta \rho_{\kvec''\sigma''\kvec'\sigma'}.
\label{drpp}
\eeq 
Then the $hh$ and $pp$ matrix elements are quadratic in the $ph$
and $hp$ $\delta \rho$ matrix elements. Therefore in Eq.~(\ref{Trhf})
the term  $Tr[{h}_{0}\delta \rho]=\sum_{\kvec} \epsilon_{\kvec} \delta
\rho_{\kvec\kvec}$ is first order in the $hh$ and $pp$ matrix elements but yields a quadratic contribution in the $ph$ and $hp$
 matrix elements. The density deviations that are off-diagonal
in the spin index contribute to the magnetic
susceptibility\cite{sei04b} but not to the charge susceptibility,
therefore in the following they will be neglected. 
One obtains 
\begin{eqnarray}
&&Tr[{h}_{0}\delta \rho] = \sum_{k>k_F , \sigma}\epsilon_{\kvec\sigma}\delta \rho_{\kvec\kvec\sigma} + \sum_{k' <k_F , \sigma}\epsilon_{\kvec'\sigma}\delta \rho_{\kvec'\kvec'\sigma} \nonumber \\ &=& \sum_{k>k_F , k'<k_F ; \sigma}(\epsilon_{\kvec\sigma}-\epsilon_{\kvec'\sigma})\delta \rho_{\kvec\kvec'\sigma}\delta \rho_{\kvec'\kvec\sigma}
\label{trhg}
\end{eqnarray} 
In the GA the interacting dispersion $\epsilon_{\kvec\sigma}$ is related to the bare dispersion $\epsilon^0_{\kvec\sigma}$ through the relation $\epsilon_{\kvec\sigma} =z^2_0 \epsilon^0_{\kvec\sigma}$.
Eq.~(\ref{trhg}) shows that the first nonzero contribution beyond the
saddle point energy is of second-order in the particle-hole density
deviations which are our independent variables.

We proceed by considering $\delta E_{e}^{(2)}$ separately for $n=1$ and general $n$.

\subsection{Half-filling case}
Closed formulas can be obtained at half-filling which illustrate in a
simple manner the physics. This generalizes the computation done by
Vollhardt\cite{vol84} to arbitrary momenta.
The second order energy contribution for the local charge deviations is 
\beqa
\delta E_{e}^{C(2)} &=&\frac{1}{2N} \sum_{\qvec} V_\qvec \delta \rho_{\qvec}\delta \rho_{-\qvec}\nonumber +\frac{1}{N}\sum_{\qvec}L_\qvec \delta \rho_{\qvec}\delta D_{-\qvec}\\ &+&\frac{1}{2N}\sum_{\qvec} U_\qvec \delta D_{\qvec} \delta D_{-\qvec}
\label{Erdhf}
\eeqa being $\delta D_\qvec$=$\dfrac{1}{N} \sum_{i}e^{-i\qvec\cdot \rvec_i}\delta D_i$ and 
\begin{eqnarray}
V_\qvec &=&  \frac{e^0 z_0}{2}(z_{++}''+ 2z_{+-}''+z_{--}'') + \frac{2(z^{'})^2}{N}\sum_{\kvec\sigma}\epsilon^{0}_{\kvec + \qvec, \sigma} n_{\kvec\sigma}\nonumber \\
L_\qvec &=&  2e^{0}z_{0}z^{''}_{+D} + \frac{2z^{'}z^{'}_D}{N}\sum_{\kvec\sigma}\epsilon^{0}_{\kvec + \qvec, \sigma} n_{\kvec\sigma}\nonumber \\
U_\qvec &=&  2e^0z_0 z_{D}''
+\frac{2(z_D')^2 }{N}
\sum_{\kvec\sigma} \epsilon_{\kvec+\qvec,\sigma}^0 n_{\kvec\sigma} 
\label{LVUhf}
\end{eqnarray}
where $e^0 =
\frac{1}{N}\sum_{\kvec\sigma}\epsilon^{0}_{\kvec\sigma}n_{\kvec\sigma}$. 
$z'$ and
$z''$ denote the derivatives of the hopping factors given in the
Appendix.  
%It is convenient to explicit the labels given in Eq.~(\ref{trhg}) to
%$p$ and $h$ states. In the following if we use sums restricted to
%$k>k_F$ ($p$ states) and $k+q<k_F$ ($h$ states) or viceversa, we will
%denote them as $\tilde{\sum}$.    
Using Eq.~(\ref{hgz}) we can eliminate the $\delta D$ deviations in Eq.~(\ref{Erdhf}) so that finally the energy functional depends on $\delta \rho$ deviations alone, {\em i.e.} $\tilde{E}^{GA}_{e}[\rho]$=$E^{GA}_{e}[\rho, D(\rho)]$. 
We find  
\begin{eqnarray*}
\delta D_{\pm \qvec} = - U_{\qvec}^{-1} L_{\qvec} \delta \rho_{\pm \qvec}.
\label{GAaf}
\end{eqnarray*}
Thus the energy $\tilde{E}^{GA}_{e}[\rho]$ is 
\begin{eqnarray}
\tilde{E}_{e}^{GA}[\rho] &=& E_{e0} + \frac{1}{N}{\sum_{\kvec
    \qvec\sigma}}(\epsilon_{\kvec\sigma} -
\epsilon_{\kvec+\qvec,\sigma}) \delta
\rho_{\kvec\sigma;\kvec+\qvec,\sigma}^{ph}
\delta\rho_{\kvec+\qvec,\sigma;\kvec\sigma}^{hp} \nonumber \\ &+&
{\sum_{\qvec}}f_{-\qvec} \delta
\rho_{\qvec} +
\frac{1}{2N}{\sum_{\qvec}}A_{\qvec}\delta
\rho_{\qvec}\delta \rho_{-\qvec} 
\label{EGAhf}
\end{eqnarray}
 where 
\begin{eqnarray}
A_{\qvec}=V_\qvec - L_{\qvec}^2 U_{\qvec}^{-1} 
\label{Wqaf}
\end{eqnarray}
 is the GA residual interaction kernel. We have introduced the
 notation $\delta
\rho_{\kvec\sigma;\kvec',\sigma}^{ph}$ to indicate that only
 $ph$ elements should be taken into account i.e. sums are 
 restricted to  $k>k_F$ and $k'<k_F$.  The density deviations can be
 decomposed in $ph$, $hp$, $pp$ and $hh$ contributions.  $ph$ and
$hp$ matrix elements contribute quadratically to Eq.~\eqref{EGAhf}
 whereas $pp$ and $hh$ are higher order so
one should substitute 
\begin{eqnarray} 
\delta \rho_{\qvec} = \frac{1}{N}\sum_{\kvec\sigma}\delta \rho_{\kvec+\qvec,\kvec\sigma}^{ph}+\frac{1}{N}\sum_{\kvec\sigma}\delta \rho_{\kvec+\qvec,\kvec\sigma}^{hp}
\label{drqkph}
\end{eqnarray}

Minimizing Eq.~(\ref{EGAhf}) with respect to the deviations $\delta
\rho$ and considering the constraints on the momenta one finds the following equation for $\delta \rho_\qvec$ 
\begin{equation}
\delta \rho_{\qvec}=
-\chi^{0}_{\qvec}f_{\qvec} - \chi^{0}_{\qvec}A_{\qvec} \delta
\rho_{\qvec}. 
\label{mingz}
\end{equation} 
Here $\chi^{0}_{\qvec}$ is the static Lindhard function, that is the charge susceptibility of the non interacting {\em quasiparticles} 
\begin{eqnarray}
\chi^{0}_{\qvec} = - \frac{1}{N}\sum_{\kvec\sigma}\frac{n_{\kvec+\qvec,\sigma}-n_{\kvec\sigma}}{\epsilon_{\kvec+\qvec,\sigma}-\epsilon_{\kvec\sigma}}
\label{Li0}
\end{eqnarray} 
Notice that $\epsilon_{\kvec\sigma}$ is renormalized by interactions. 
From Eq.~(\ref{mingz}), we obtain the linear response equation \cite{rin80, Noz90}
\begin{eqnarray}
\delta \rho_{\qvec} = - \kappa_{\qvec}f_{\qvec}
\label{LIRE}
\end{eqnarray} 
with the GA+RPA static response function
\begin{eqnarray}
\kappa_{\qvec} = \frac{\chi^{0}_{\qvec}}{1+\chi^{0}_{\qvec}A_{\qvec}}
\label{kp}
\end{eqnarray}
$\kappa_{\qvec=0}$ is the charge compressibility studied by Vollhardt
for $n=1$.\cite{vol84} 
Putting $A_{\qvec}$=0 and $z_0 =1$, one recovers $\kappa_{\qvec}$=$\chi^{00}_{\qvec}$ for  non-interacting electrons.

In the case of nearest neighbor hopping on a $d-$dimensional cubic lattice
we use the relation 
\begin{eqnarray*}
\frac{1}{N}\sum_{\kvec\sigma}\epsilon^{0}_{\kvec + \qvec, \sigma} n_{\kvec\sigma} = \frac{e^{0}}{d}\sum_{\nu=1}^d\cos q_{\nu} = e^{0}\eta_{\qvec}
\label{E0e0}
\end{eqnarray*}where 
\begin{equation}
\eta_\qvec = \frac{1}{d}\sum_{\nu=1}^d \cos q_\nu
\label{dcos}
\end{equation} 
with $\nu=1,...,d$ according to the dimension $d$.  
Using Eqs.~(\ref{LVUhf}) and (\ref{Wqaf}), $A_\qvec$ takes the following form 
\beqa
\frac{A_{\qvec}}{e^0}&=& \frac{z_{0}}{2}(z_{++}''+ 2z_{+-}''+z_{--}'') + 2(z^{'})^2 \eta_{\qvec} \nonumber \\ &-& 2\frac{(\eta_{\qvec} z' z'_{D} + z_0 z''_{+D})^2}{\eta_{\qvec} (z'_{D})^2 + z_0 z''_{D}}
\label{Wqe0}
\eeqa 
From Eq.~(\ref{Wqe0}) we find an analytical expression of the effective interaction $A_\qvec$ in term of the Coulomb interaction $U$ for $U < U_c$ %(Fig.~\ref{Wq1Dn1}) 
\beqa
A_\qvec = \frac{U(U_c + U)(U-2U_c)}{4U_c(U -U_c)}
\label{WqU}
\eeqa being $U_c = 8 |e^0_{n=1}|$.
Thus we find that at half-filling $A_\qvec$ is independent from the momentum $\qvec$. In the weak coupling limit, we recover the HF-RPA result  $A_\qvec \approx U/2$. $A_\qvec$ is an increasing function of $U$, diverging at $U=U_c$ 
\beqa
\lim_{U\rightarrow U_c^{-}} A_\qvec = \frac{U^{2}_c}{2(U_c -U)}
\label{WqUc}
\eeqa 
Then at the Mott transition not only the charge compressibility vanishes,~\cite{BR70} but also the susceptibility $\kappa_\qvec$ for any momentum [see Eq.~(\ref{kp})]. 

\subsection{Arbitrary filling case}
\label{anyfil}

The full derivation of $\delta E_{e}^{(2)}$ is given in the Appendix, both in real and momentum space. 

In addition to $\delta \rho_\qvec$, we find that it is convenient to
introduce the quantity 
$$\delta T_i=\sum_{j\sigma} t_{ij}(\delta
\rho_{ji\sigma} + \delta \rho_{ij\sigma})$$
 and its Fourier transform 
\beq
\delta T_{\qvec}=\sum_{\kvec\sigma}(\epsilon_{\kvec\sigma}^{0} + \epsilon_{\kvec+\qvec,\sigma}^{0})\delta \rho_{\kvec+\qvec,\sigma; \kvec\sigma}
\label{Tq}
\eeq    
$\delta T_{\qvec}$ corresponds to intersite charge fluctuations, while
$\delta \rho_{\qvec}$ describes the local ones.  

The second order-energy expansion for the charge deviations is given by 
\begin{eqnarray}
\delta E_{e}^{C(2)} &=& \frac{1}{N}\left[ \frac{1}{2}\sum_{\qvec} V_\qvec \delta \rho_\qvec \delta \rho_{-\qvec}
+ z_0 z_{D}'\sum_{\qvec} \delta D_\qvec \delta T_{-\qvec}\right. \nonumber \\
&+& \frac{1}{2} z_0(z'+z_{+-}')\sum_{\qvec} \delta T_\qvec \delta \rho_{-\qvec}
 \nonumber \\
&+&\left.  \sum_\qvec L_\qvec \delta \rho_\qvec \delta D_{-\qvec}
+ \frac{1}{2} \sum_\qvec U_\qvec \delta D_\qvec \delta D_{-\qvec}\right]
\label{eskspace}
\end{eqnarray}
with the following definitions 
\begin{eqnarray}
V_\qvec &=& \frac{e^0 z_0}{2}(z_{++}''+ 2z_{+-}''+z_{--}'')
+\frac{(z'+z_{+-}')^2}{2N}
\sum_{\kvec\sigma} \epsilon_{\kvec+\qvec,\sigma}^0n_{\kvec\sigma} \nonumber \\
L_\qvec &=& e^0z_0 (z_{+D}''+ z_{-D}'')
+\frac{z_D'(z'+z_{+-}')}{N}
\sum_{\kvec\sigma} \epsilon_{\kvec+\qvec,\sigma}^0n_{\kvec\sigma} \nonumber \\
U_\qvec &=& 2e^0z_0 z_{D}''
+\frac{2(z_D')^2}{N}
\sum_{\kvec\sigma} \epsilon_{\kvec+\qvec,\sigma}^0 n_{\kvec\sigma} \label{eq:vlu}
\end{eqnarray}where $z'$ and $z''$ denote derivatives of the hopping factors which are given in the Appendix.

Using Eq.~(\ref{hgz}) one can eliminate the
double occupancy deviations and arrive at the following functional
which only depends on the local and intersite charge deviations 
\begin{equation}\label{eq:eexp}
\delta E_{e}^{C(2)} =\frac{1}{2N}\sum_{\qvec} \left(\begin{array}{c}
\delta \rho_\qvec \\ \delta T_{\qvec}\end{array}\right)
\left(\begin{array}{cc} A_\qvec & B_\qvec \\
B_\qvec & C_\qvec \\
\end{array}\right)
\left(\begin{array}{c}
\delta \rho_{-\qvec} \\ \delta T_{-\qvec}\end{array}\right) 
\end{equation}where 
\beqa
W_{\qvec} =\left(\begin{array}{cc}
A_{\qvec} & B_{\qvec}\\
B_{\qvec} & C_{\qvec}\end{array}\right)
\label{Wqgen}
\eeqa 
is the interaction kernel. 
The elements of $W_{\qvec}$ are given by
\begin{eqnarray*}
A_\qvec &=& V_\qvec -\frac{L_\qvec^2}{U_\qvec} \\
B_\qvec &=& %\frac{1}{2}\left\lbrack 
z_0 (z'+z_{+-}')- 2 z_0 z'_D
\frac{L_\qvec}{U_\qvec} \\ % \right\rbrack 
C_\qvec &=& - \ \  \frac{(z_0z_D')^2}{U_\qvec}
\end{eqnarray*}
Since the energy expansion in Eq.~(\ref{eq:eexp}) is a quadratic form in
$\delta \rho_{\qvec}$ and $\delta T_\qvec$ (see also Eq.(\ref{Tq})), it is useful to introduce
the following representation for the static Lindhard function $\chi^{0}_{\qvec}$ 
\begin{widetext}
\begin{equation}
\label{sus}
\chi^{0}_{\qvec}= - \frac{1}{N} \sum_{\kvec\sigma}\left(\begin{array}{cc}
1 & \epsilon^0_{\kvec\sigma} + \epsilon^0_{\kvec+\qvec,\sigma}\\
\epsilon^0_{\kvec\sigma} + \epsilon^0_{\kvec+\qvec,\sigma} & (\epsilon^0_{\kvec\sigma} + \epsilon^0_{\kvec+\qvec,\sigma})^2 \end{array}\right)
 \frac{n_{\kvec+\qvec,\sigma} - n_{\kvec\sigma}}
{\epsilon_{\kvec+\qvec,\sigma}-\epsilon_{\kvec\sigma}}.
\end{equation}
\end{widetext}
The calculations proceed analogously to the half-filling case. We find that the RPA series for the charge excitations then corresponds to the following Bethe-Salpeter equation 
\begin{equation}
\label{eqrpa}
\chi_{\qvec} = \chi_\qvec^{0} - 
\chi_\qvec^{0} W_\qvec \chi_\qvec
\end{equation}
For general fillings the response function $\chi_\qvec$ is given by a $2\times 2$ matrix whose element $(\chi_\qvec)_{11}$ is the charge susceptibility $\kappa_\qvec$.

For $U=\infty$ we can derive an analytical expression for $A_\qvec$ valid for any filling $n<1$ and any dimension 
\beqa
A_\qvec &=& (-e^0) \frac{5-4n-\eta_\qvec}{(1-n)(2-n)^3} 
\label{WGe}
\eeqa
We notice for later use that, 
since $e_0<0$, the interaction has a minimum at $\qvec=0$, a maximum
at  $\qvec=(\pi,\pi,...)$ and  diverges as $n\rightarrow 1$.  
Within the GA+RPA the charge vertex $\Gamma_\qvec$,\cite{Kel95, Hua03, Koc04, Cap04, Cit01, Cit05} 
introduced in Section II, is
\begin{eqnarray}
\Gamma_\qvec  = \beta \frac{(\chi_\qvec)_{11}}{(\chi^0_\qvec)_{11}}.
\label{Gamma}
\end{eqnarray}  
We stress that $\Gamma_\qvec$ is the renormalized {\em quasiparticle}-$ph$ coupling.  
The renormalized coupling for the {\em electrons} instead 
is $g_\qvec = Z_\qvec \Gamma_\qvec$, with $Z_\qvec$ 
the quasiparticle weight, given by $z_0^2$ in the GA. 
In the following computations for $\Gamma_\qvec$ and $g_\qvec$, we will use $\beta=1$.

In Sections IV  and V we will describe the behaviour of the charge susceptibility
 $\kappa_\qvec \equiv (\chi_\qvec)_{11}$ obtained using Eq.~(\ref{eqrpa}). First we 
test the performance of the GA+RPA approach in  $d=\infty$: this is obviously
the most suitable case for the GA+RPA formalism particularly because the GA 
corresponds to the exact evaluation of the Gutzwiller variational wave-function
in this limit. We then will move to the 2d case, which is physically relevant for quasi 
2d-materials like cuprates.

In the following sections we present results for hopping restricted to
nearest neighbor $t_{ij}=-t$ and put $t=1$ which makes the energy and  
the charge susceptibilities dimensionless.  Occasionally we will
explicitly rescale the interaction by the Brinkman-Rice transition
$U_c$ leaving the compressibility units untouched.  

\section{Results in infinite dimensions}
We consider the case of a hypercubic lattice in infinite dimansions with nearest-neighbor 
hopping, where the density of states per spin is given by 
$$
\rho^0(\omega)=\sqrt{\frac{2}{\pi}}\frac{1}{t}\exp\left(-\frac{\omega^2}{2t^2}\right) .
$$
In this case a momentum dependence in the response is still present via the quantity
 \cite{mul89} 
\begin{eqnarray*}
\eta_\qvec = \frac{1}{d}\sum_{\nu=1}^d \cos q_\nu
\end{eqnarray*}
which enters the interaction kernel $W_\qvec$ and the correlation functions $\chi_\qvec$.
For example, the non interacting susceptibility reads
\begin{eqnarray*}
{\bf \chi}_\qvec^0&=&-4\int_{-\infty}^{\mu}d\omega'\int_{\mu}^{\infty}
d\omega'' \left(\begin{array}{cc}
1 & \frac{\omega'+\omega''}{z_0^2} \\
\frac{\omega'+\omega''}{z_0^2} & \frac{(\omega'+\omega'')^2}{z_0^4}
\end{array}\right) \\
&\times&  \frac{\Lambda_\qvec(\omega',\omega'')}{\omega'-\omega''}
\end{eqnarray*}
with
\begin{eqnarray*}
&&\Lambda_\qvec(\omega',\omega'')=\frac{1}{2\pi z_0^4 t^2\sqrt{1-\eta_\qvec^2}}\times\\
&&\exp\left\lbrack -\frac{1}{4z_0^4t^2}\left\lbrace 
\frac{(\omega'-\omega'')^2}{1-\eta_\qvec} + \frac{(\omega'+\omega'')^2}{1+\eta_\qvec}
\right\rbrace \right\rbrack
\end{eqnarray*}
and in the two limiting cases $\eta_{\qvec=0}=1$ and $\eta_{\qvec=\QVEC}=-1$ one can give 
analytical expressions for the static susceptibility matrices of the Lindhard function 
\begin{eqnarray*}
{\bf \chi}_0^0 = 2 \rho(\mu)\left( \begin{array}{cc}
1 & 2 \mu \\
2 \mu & 4 (\mu)^2 \end{array} \right) \\
\chi_\QVEC^0 =
\frac{1}{\sqrt{2\pi}z_0^2t}\left( \begin{array}{cc}
1 & 0 \\
0 & 0 \end{array} \right) E_1\left(\frac{\mu^2}{2t^2z_0^4}\right),
\end{eqnarray*}
where $\QVEC$ is the momentum $(\pm \pi, \pm \pi, \pm \pi...)$ and $E_1(x)$ denotes 
the exponential integral.~\cite{abr65}

\begin{figure}[t]
\includegraphics[width=8cm,clip=true]{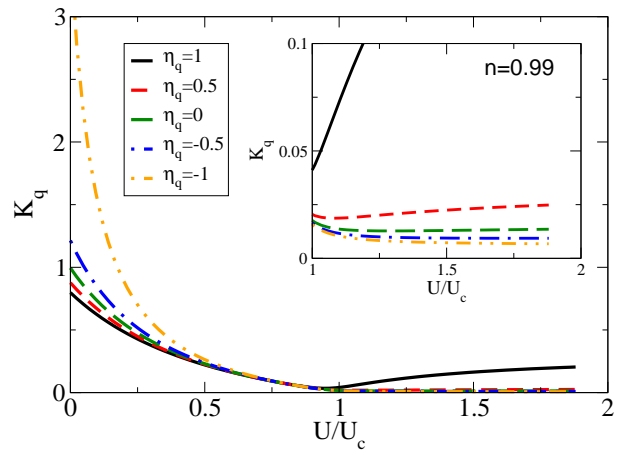}
\caption{$d=\infty$ - Charge susceptibility as a function of $U/U_c$
  for $n=0.99$. The inset shows an enlargement of the high $U$ region. }
\label{fig3}
\end{figure}
\begin{figure}[t]
\includegraphics[width=8cm,clip=true]{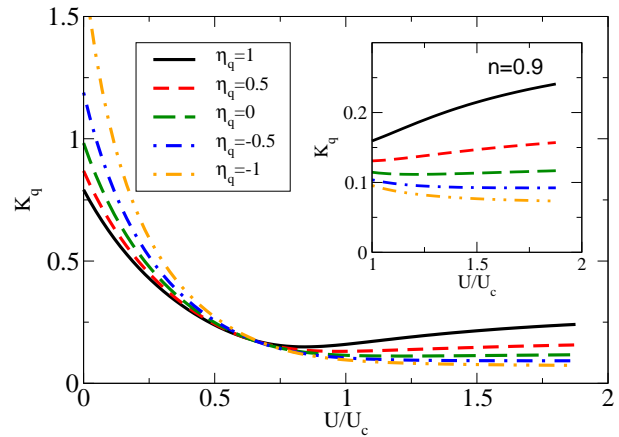}
\caption{$d=\infty$ - Charge susceptibility as a function of $U/U_c$ for $n=0.9$.}
\label{fig4}
\end{figure}
\begin{figure}[t]
\includegraphics[width=8cm,clip=true]{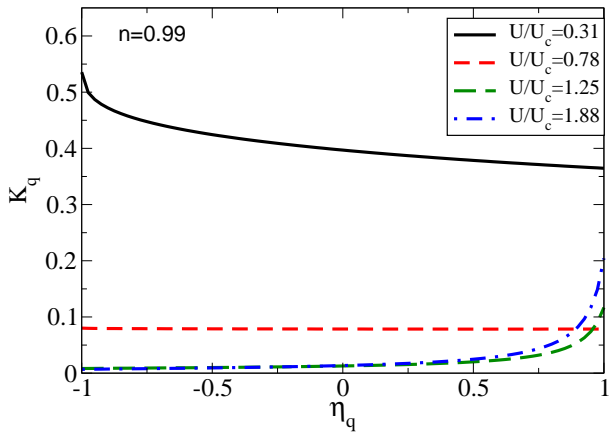}
\caption{$d=\infty$ - Charge susceptibility as a function of
  $\eta_\qvec$ for $n=0.99$ for various values of the interaction
strength $U/U_c$.}
\label{fig1}
\end{figure}
\begin{figure}[t]
\includegraphics[width=8cm,clip=true]{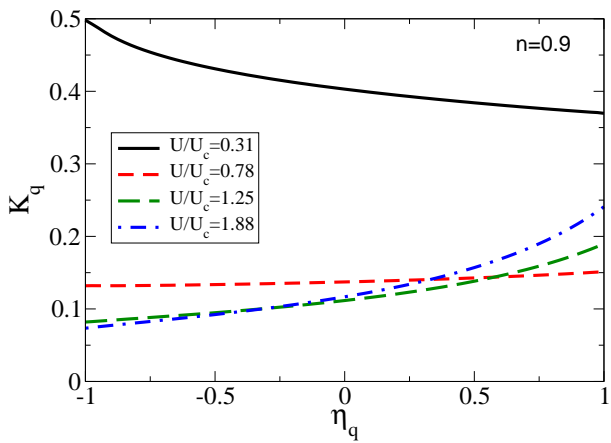}
\caption{$d=\infty$ - Charge susceptibility as a function of $\eta_\qvec$ for $n=0.9$ 
for various values of the interaction
strength $U/U_c$.}
\label{fig2}
\end{figure}

As discussed in Ref.~\onlinecite{mul89}, for a generic $\qvec$ the corresponding $\eta_\qvec$ is 
trivially zero; only for special $\qvec$ $\eta_\qvec$ takes values between -1 and +1. 
In particular 
for the hypercubic lattice the relevant $\qvec$'s are the ones along the diagonal 
$(0,0,0...) - (\pi,\pi,\pi...)$ (and the other equivalent directions of the hypercubic lattice, 
which form a set of measure zero).
This means that for this infinite dimensional lattice, the study of the momentum depencence
 of the quantities is sensitive to effects in the (1,1,1...) direction of the Brillouin zone,
while it cannot access the other directions like, for instance (1,0,0,0...) or (...0,0,0,1). 

In Figs.~\ref{fig3} and \ref{fig4} we show the dependence on $U$ for
selected momenta at $n=0.99$ (almost half-filling)  and $n=0.9$ respectively. 
The curves correspond to different values of the interaction strength 
 in units of the critical $U_c = 8 \sqrt{2/\pi} t$, which is the interaction at which
in the infinite dimensional hypercubic lattice electrons undergo the metal-insulator 
transition at half-filling ($n=1$) in the GA. At small $U$
the susceptibility has a strong enhancement at $\qvec_c=\QVEC$, this
is due to the nesting of the lattice and leads to the Peierls CDW
instability in the presence of coupling to the lattice
[Eq.~\eqref{eq:lambdac}]. 

Starting from the small-$U$ side, the charge susceptibility is
suppressed upon approaching $U_c$ and then slightly increases again when $U$ is
further increased.
The behavior of $\kappa_\qvec$ for $U>U_c$ strongly depends on momentum
and the suppression is most effective for large $q$ and close to
half-filling.

Perfect nesting occurs only at half-filling due to the matching of the
Fermi (hyper)surface when translated by $\QVEC$. In analogy with low
dimensional systems one may wonder whether 
away from half-filing an incommensurate CDW is favored.

Figs.~\ref{fig1} and \ref{fig2} show the momentum dependence
of the susceptibility for density $n=0.99$ and $n=0.9$ and
various values of $U/U_c$, respectively.

At small $U$ one finds the nesting-induced enhancement 
at $\qvec=\QVEC$ for both fillings indicating that 
incommensurate CDW formation is not favored. i.e.
there is no  shift of $\qvec_c$.

Interestingly 
at large $U$ another instability enters into play since in
this limit $\kappa_\qvec$ acquires a maximum at $\eta_\qvec=1$.
One qualitatively recovers
a momentum structure similar to what is obtained within 
the large-N expansion of the Hubbard model for $U = \infty$ \cite{Koc01,Sei00}. In this
case the residual repulsion between quasiparticles is most effective
at large momenta leading to a suppression of $\kappa_\qvec$ for $\eta_\qvec\to -1$. 
Inclusion of a Holstein coupling would induce a PS
instability in this limit.
The momentum dependence of the susceptibility becomes weak
for intermediate values of $U$, slightly below $U_c$.
All these features are most pronounced upon approaching half-filling.

The fact that for small $U/U_c$ the instability momentum is pinned at
$\QVEC$ is specific to a high-dimensional system,
where the effects of nesting of the Fermi surface are weak and the effects of doping
in changing the Fermi surface are negligible. We will see in the 
next section that this is not the case in 2d, where upon doping
$\qvec_c$ moves away from the (1,1) direction and shifts along the (1,0) direction. 

From Figs.~\ref{fig1} and \ref{fig2} we conclude that 
when $U$ is increased beyond a value of about $0.78U_c$ the maximum in the 
charge response moves from large to small momenta. This signals that
the inclusion of (momentum-independent) phonons would drive the system
towards a PS instability at large $U$'s, 
while at small $U$'s the systems would undergo a transition to a CDW state.
The behavior of the charge response also allows to infer that the $e$-$e$
correlations  suppress more severely the $e$-$ph$ coupling at large
momentum transfer than at small transferred momenta. This is
the reason why, upon increasing 
$U$, the system will undergo more easily a low-momentum instability
(PS), rather than becoming unstable at finite (and usually large) momenta.

\section{Results in two dimensions}
\label{sec:res2d}
We now move to the 2d case, which is relevant for many layered materials.
We typically worked with a $100 \times 100$ lattice for our computations. 

\begin{figure}[t]
\centering
\includegraphics[width=8.6 cm]{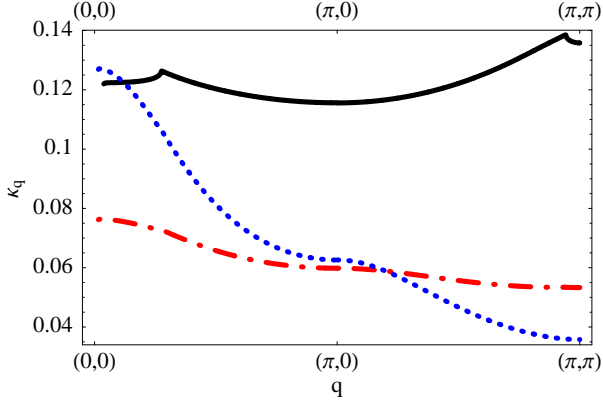} 
\caption{2d-Charge susceptibility for $n=0.9$ along the open path $\qvec = (0,0) - (\pi,0) - 
(\pi,\pi)$ for $U/U_c = 0.5$ (solid line), $U/U_c = 0.9$ (dot-dashed line) and $U/U_c = 2$ 
(dotted line).}
\label{fig:2Dkq}
\end{figure}

We start by characterizing the Peierls instability in 2d. For small  $U$ and
$n=1$ the charge susceptibility has the Peierls  
peak at $\qvec =(\pi,\pi)$, associated with the Fermi surface nesting. 
For the doped 
system the response exhibits a peak for a  $\qvec_{c}$ close to 
$(\pi,\pi)$, featuring the tendency to develop an incommensurate CDW
in the presence of phonons (Fig.~\ref{fig:2Dkq}, full line). 
The momentum of the instability undergoes a shift in the  $(1,0)$ 
direction of the Brillouin zone.  For small $U$ the peak is located at 
$\qvec_{c}=(\alpha (\delta) \pi,\pi)$
with 
$$\alpha (\delta) = 1 -0.46\delta - 1.30\delta^2...\sim
1-\delta/2.$$ This depends little on $U$ at weak coupling  due to the weak
$\qvec$-dependency of $W_\qvec$ close to $n=1$.  A behaviour
compatible to ours has been found  
also for the 2d-Holstein model\cite{Vek92} with QMC and RPA calculations
on a $8\times8$  lattice.

\begin{figure}[tp]
\centering
\includegraphics[width=8cm]{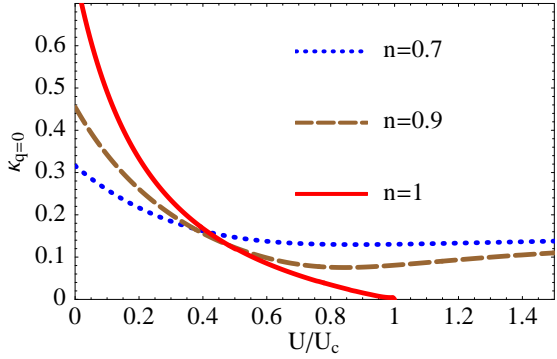}
\caption{2d-Charge compressibility as a function of $U/U_c$ for various fillings.}
\label{fig:kq0du2d}
\end{figure}

Along the (1,0) direction 
$\kappa_\qvec$ exhibits another peak at $\qvec'_c$  (the peak 
close to $\qvec=(0,0)$, full line in  Fig.~\ref{fig:2Dkq}).
This corresponds to the scattering 
between states at the (rounded) corners of the Fermi surface in
adjacent Brillouin zones. Upon increasing $U$ (cf. Fig.~\ref{fig:2Dkq}), 
the nesting induced peak structure gets lost. Simultaneously the
response at large wave-vectors becomes suppressed
and overcome by that at $\qvec=(0,0)$. This indicates  that
the order of the instabilities is reversed like in the $d=\infty$ case: 
For large $U$ the
system phase separates before the CDW instability arises.  This
behavior will become more clear upon analyzing the charge
susceptibility as a function of filling and interaction. 
In Fig.~\ref{fig:kq0du2d} we show the charge compressibility in 2d 
as a function of $U/U_c$ with $U_c = 128t/\pi^2$. 
For $U=0$ the compressibility  
is given by the  noninteracting density of states at the Fermi energy. The
latter diverges for $n\rightarrow 1$ due to the Van Hove
singularity. For $U$ close to $U_c$ one recovers a similar
behavior as in $d=\infty$. The compressibility vanishes at the Mott transition point
and has a minimum close to $U_c$ for $n\ne 1$.   
\begin{figure}[htbp]
\centering
\includegraphics[width=8.cm]{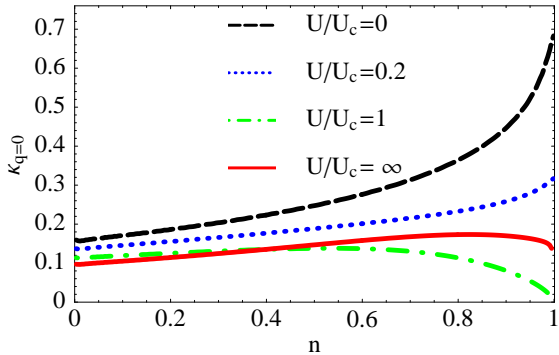}
\caption{2d-Charge compressibility as a function of $n$ for various $U/U_c$.}
\label{2dk0n}
\end{figure}

In Fig.~\ref{2dk0n} we present the compressibility as a function of $n$ for different 
values of $U/U_c$. For $n\rightarrow 0$ the compressibility can be
computed exactly using a low density expansion\cite{fab91}. The ground state
energy reads 
$$ 
\frac{E}N=\pi t n^2 -2\pi t \frac{n^2}{\ln(n)}+...
$$
where the second term is the leading correction due to interactions. 
Computing the compressibility as $\kappa_{\qvec=0}^{-1}=\partial^2 E/\partial n^2/N$
one finds that the zero density limit is given by the noninteracting
compressibility 
$$\kappa_{\qvec=0}(n \rightarrow 0)=(2\pi t)^{-1}.$$
In GA+RPA we find instead a small suppression of the zero density compressibility
with interaction. This is not surprising since 
RPA is expected to break down in the low density limit. Still this
dependence is quite small and we expect that our results are accurate at
moderate densities.  

For small $U$ the compressibility is an increasing function 
of $n$ and reaches the maximum value for $n=1$ as a consequence of the van Hove enhancement. 
For $U=U_c$,  $\kappa_{\qvec=0}$ goes to zero for $n=1$; for larger $U>U_c$, $\kappa_{\qvec=0}$ 
flattens, still exhibiting a smooth maximum for finite doping. Therefore the GA 
compressibility has a jump discontinuity for $n=1$ and $U>U_c$: its left and right 
limits are finite, while it vanishes at $n=1$.~\cite{note1} 

The qualitatively 
different behaviour of the compressibility for small $U$ and large $U$
is clear:   
for low fillings the system is weakly affected by $e$-$e$ interactions and its 
compressibility increases with $n$ no matter how large $U$ is. Approaching
half-filling the correlated nature of the system becomes relevant 
and reduces the compressibility of the electron liquid around $n=1$. 
One should also keep in mind that close to half-filling AF correlations will
become relevant. 

\begin{figure}[b]
\centering
\includegraphics[width=8. cm]{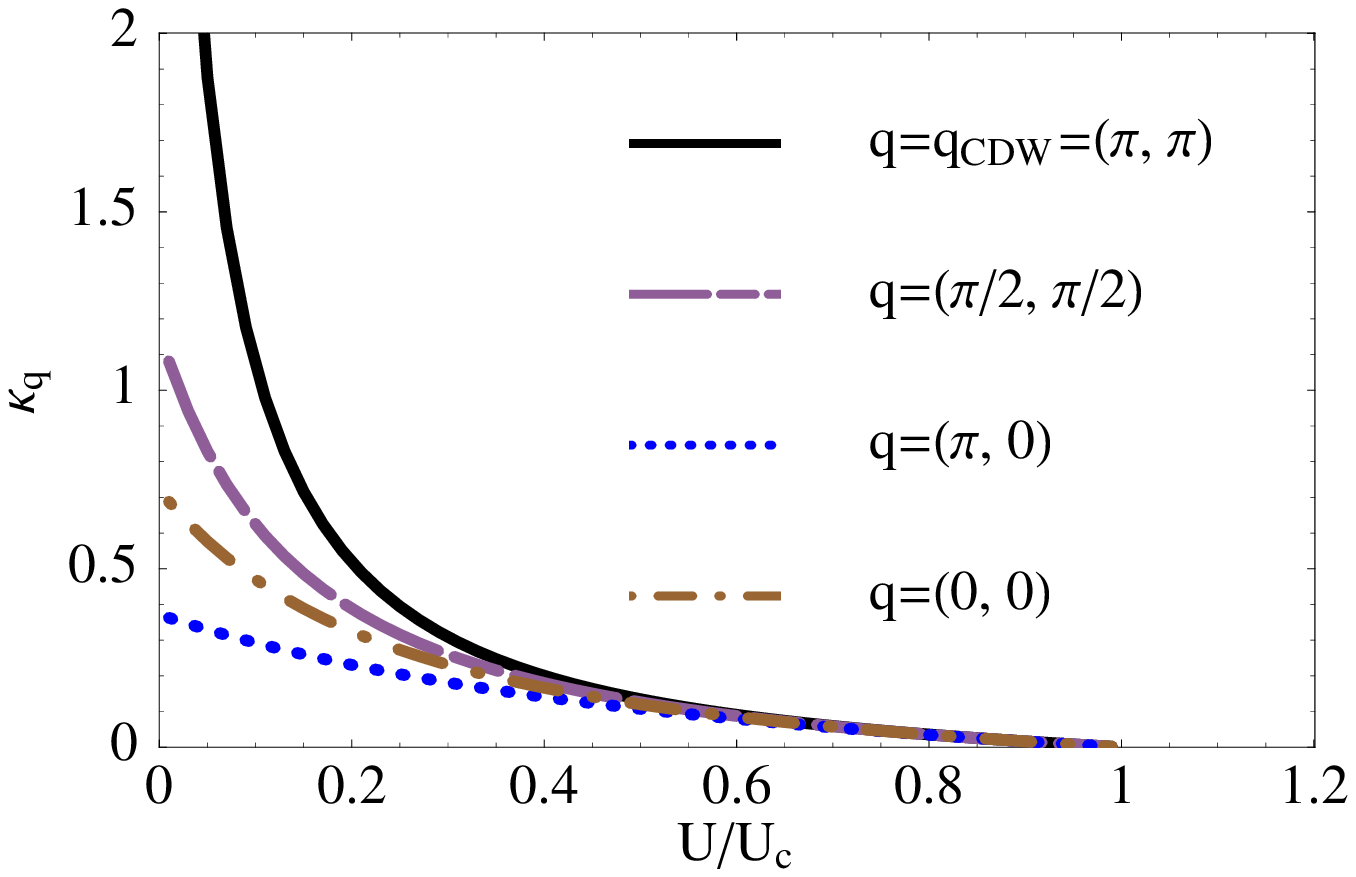}\hspace{.1 cm}
\includegraphics[width=8. cm]{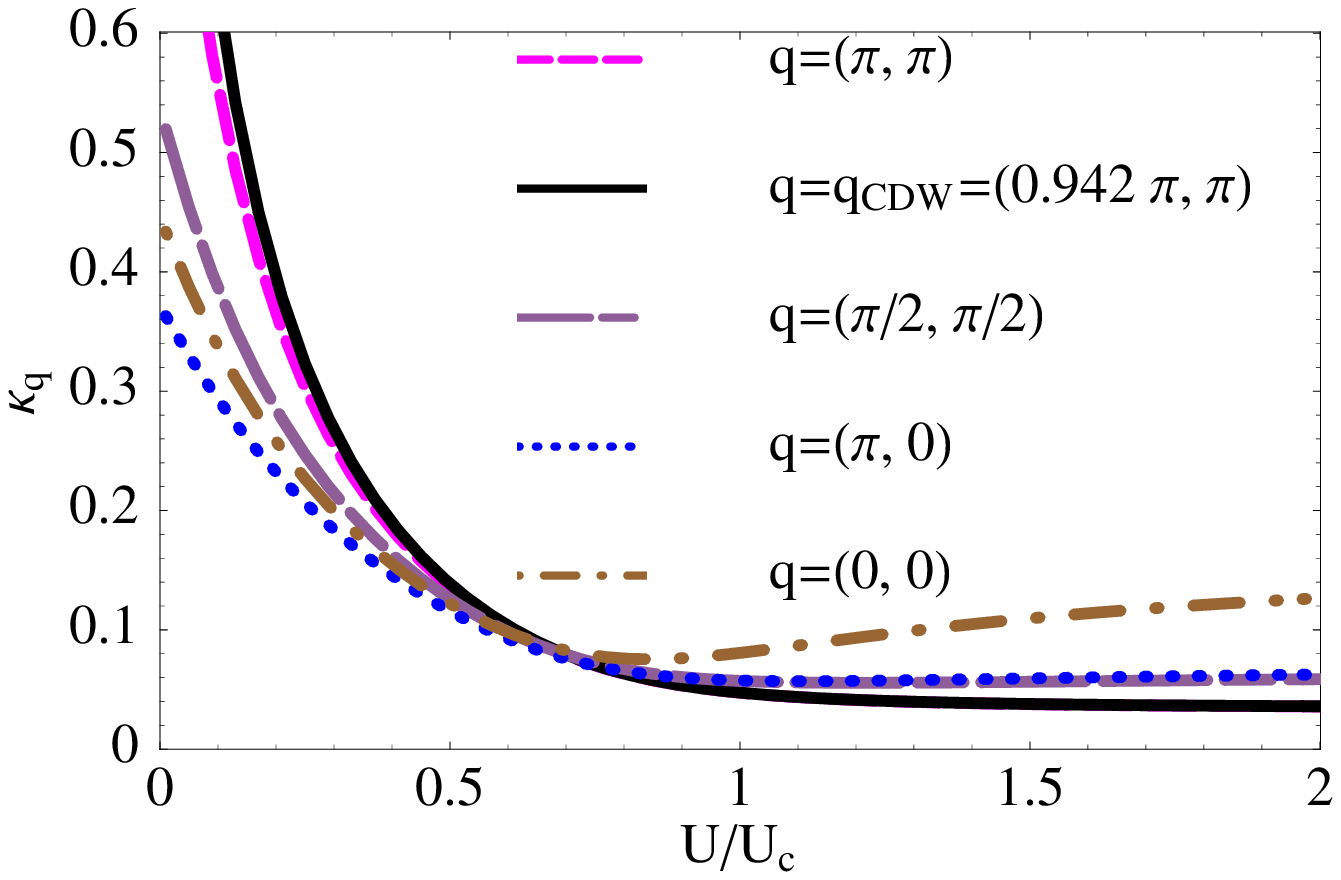}
\caption{2d-Charge susceptibility as a function of $U/U_c$ for $n=1$ (top) and $n=0.9$ (bottom).}
\label{fig:2DkU}
\end{figure}

In Fig.~\ref{fig:2DkU} we show the charge susceptibility $\kappa_\qvec$ as a 
function of $U/U_c$ for $n=1$ and $n=0.9$ and selected momenta. 
At small momenta the charge
susceptibility  
is close to the compressibility in both cases. As the momentum
approaches $\qvec=\qvec_c$  
the charge susceptibility takes its highest values for small $U$. In
particular for $n=1$  (perfect nesting) and $U=0$ the charge susceptibility diverges
indicating that an infinitesimal $\lambda$ renders the system
unstable. The susceptibility, however, is strongly suppressed by $U$ and at half-filling for 
any momentum goes to zero for $U=U_c$. Therefore, as for $d=\infty$,
$e$-$e$ interactions renormalize   
the noninteracting  CDW instability which thus needs a finite
$\lambda$ to occur.    

At small doping 
$\kappa_\qvec$ is finite and shows a shallow minimum close to $U=U_c$. As in $d=\infty$
we see from the larger $\kappa_\qvec$ that for $U \gtrsim U_c$  the PS
instability becomes dominant. 

The dominance of the PS instability at
large $U$ can be understood from the strong coupling results close to
$n=1$. In this case we expect  Eq.~(\ref{kp}), derived for $n=1$,  to be a good
approximation. For  $U \gg U_c$, $\kappa^0_\qvec A_\qvec$ takes large values; then, using Eq.~(\ref{kp}), 
we find simply that the compressibility saturates as a function of $U$
at a value $\kappa_\qvec \approx 1/A_\qvec$ 
giving a maximum in $\kappa_\qvec$ at $\qvec=0$ (cf. Eq.~\eqref{WGe}).  
Our results are consistent with Ref.~\onlinecite{CDCG95} where C. Castellani {\em et al.} 
find PS in a $U=\infty$ slave boson (SB) investigation at $T=0$. 
On the other hand, SB calculations for 2-d systems \cite{Koc04,Cap04} 
have found PS for $T \approx t$  
even in the absence of phonons, while for lower $T$ an homogeneous
state is preferred again. 
This reentrant behavior has not been confirmed by other techniques and
should be taken with care due to the poor performance of mean-field
 SB techniques at finite $T$.

\begin{figure}[tbp]
\centering
\includegraphics[width=8.6 cm]{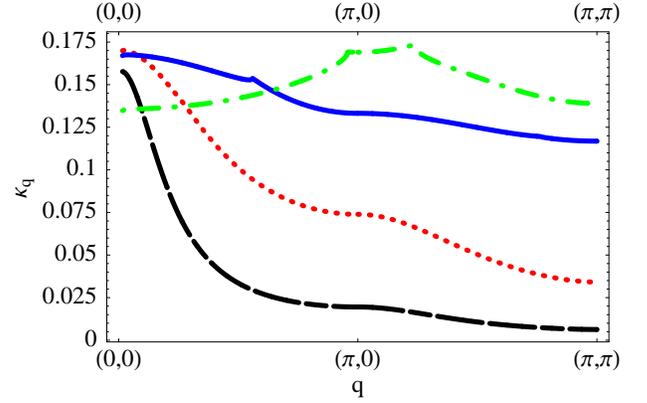}
\caption{2d-Charge susceptibility for $U/U_c = 100$ along the open path 
$\qvec = (0,0) - (\pi,0) - (\pi,\pi)$ for $n = 0.98$ (dashed line), 
$n = 0.9$ (dotted line), $n =0.7$ (solid line) and $n =0.4$ (dotted-dashed line).}
\label{fig:2DkqUINF}
\end{figure}

To illustrate the strong coupling behavior,  
in Fig.~\ref{fig:2DkqUINF} we show the charge susceptibility for $U/U_c = 100$ along the 
open path $\qvec = (0,0) - (\pi,0) - (\pi,\pi)$. 
If we consider fillings quite close to 
$n=1$, the Coulombic repulsion completely suppresses the CDW peaks which upon
doping become visible again for $n\sim0.7$. 
Clearly, more evident CDW peaks appear for 
lower fillings: for $n \ll 1$, the effects of $e$-$e$ interactions are weak even for very 
large $U$.  For $n < 0.5$, we find that the ground state is a CDW and that the susceptibility 
exhibits quite different features with respect to the fillings $n \gtrsim 0.7$. 

The most noticeable 
feature of Fig.~\ref{fig:2DkqUINF} is the peak in $\kappa_\qvec$ centered at $\qvec=(0,0)$: 
it gets narrower as the doping $\delta$ goes to zero. 
Therefore upon reducing $\delta$ the charge response to a local perturbation
spreads more and more out in space and the small ${\bf q}$ peak width is
a measure for the corresponding inverse screening length.  
We can give an analytical interpretation 
of this behaviour, using the approximate relation $\kappa_\qvec \approx 1/A_\qvec$ for $\delta 
\ll 1$ and adopting for $A_\qvec$ its $U=\infty$  form (see Sect.~\ref{anyfil}).
We have checked that for low doping  the $U=\infty$ form of $1/A_\qvec$ gives a quite accurate 
representation for the whole $\kappa_\qvec$ curve. In particular, if we consider the low $q$ 
expansion of $1/A_\qvec$, we observe that the $\qvec=(0,0)$ peak of the susceptibility is well 
fitted by a Lorentzian peak of half-width $\tilde{q} = \sqrt{8\delta}$ in the $x$- and 
$y$-directions. 
It is worth noting that the typical momentum associated to the peak depends only on 
the doping $\delta$ 
and not on the energy $e^0$. In the low $q$ limit  a relation $\tilde{q}\sim \sqrt{\delta}$ can
also be found using the $U=\infty$ single SB quasiparticle interaction.\cite{Sei00} 

\begin{figure}[t]
\centering
\includegraphics[width=7.5cm]{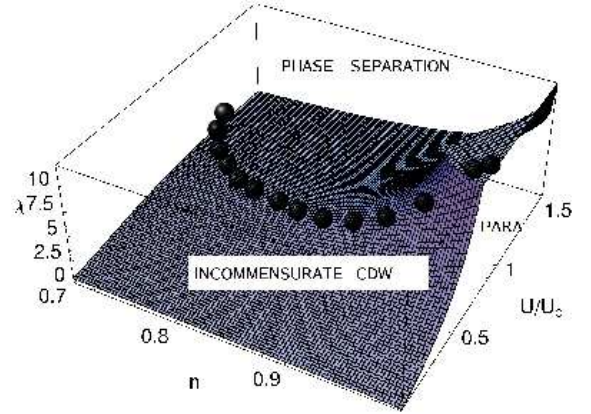}
\caption{Instability surface $\lambda(n,U/U_c)$ of the paramagnet towards CDW and PS: the 
sphered-line marks the transition between the CDW state and the state with PS.}
\label{2DPhDiaTot}
\end{figure}

All these considerations on the pure electronic response lead us to the phase diagram of the 
$e$-$ph$ system in Fig.~\ref{2DPhDiaTot}.
The main outcome is that the large $e$-$e$ interaction changes the nature of the charge
instability from an incommensurate CDW to a PS. At large $U\gtrsim U_c$ and moderate doping
this effect already occurs for small values of the bare $e$-$ph$ coupling $\lambda$. This is
the result of a compromise between a reduced quasiparticle kinetic energy (which renders the system prone
to instabilities) and the modest screening of the $e$-$ph$ coupling, when 
one is away from
the Mott-insulating phase at half-filling. The fact, that the screening is 
less important
at small transferred momenta obviously favors the occurrence of PS at $\qvec_c=0$ with respect to the incommensurate CDW.

\begin{figure}[b]
\centering
\includegraphics[width=9cm]{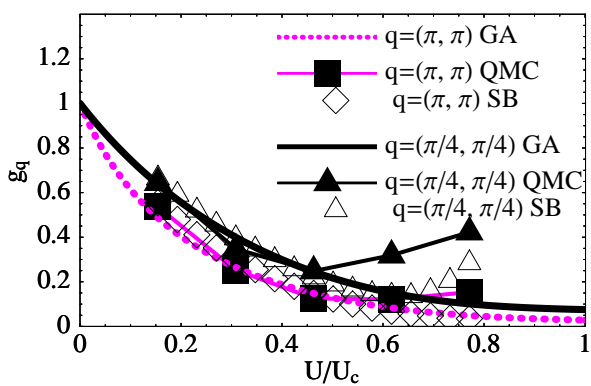}
\caption{2d-Renormalized $e$-$ph$ coupling as a  function of $U/U_c$ for $n\approx0.88$ at two different
momenta.  For comparison we report the results obtained at finite T ($T=0.5$) with 
SB \cite{Koc04} (empty symbols) and QMC \cite{Hua03} (filled symbols)  (these latters are for 
finite Matsubara electron frequency $\omega=\pi/2$).
All the results are obtained for a $8 \times 8$ 
lattice (see Fig.~3 of Ref.~\onlinecite{Koc04} and Fig.~4a of Ref.~\onlinecite{Hua03}).}
\label{2Dgq}
\end{figure}

In case of the 2-d systems we now consider explicitly the renormalized $e$-$ph$ coupling  $g_\qvec=z_0^2\Gamma_\qvec$ for the bare electrons.
In Fig.~\ref{2Dgq} we show the behaviour of the quantity $g_\qvec$ as a function of the $e$-$e$ interaction.

The trend is quite clear: the suppression of the 
bare $e$-$ph$ coupling is stronger for large $q$. We compare our results at $T=0$ 
with SB  \cite{Koc04} and QMC \cite{Hua03} calculations at $T=0.5$ (these latters are also
performed at a finite Matsubara frequency of the incoming and outgoing fermions 
$\omega = \pi T$). The agreement is generically quite good. However, in the 
finite-T results an upturn of $g_\qvec$ 
is also present (more pronounced for small $\qvec$), which was interpreted in Ref. \onlinecite{Koc04} 
as the signature of an incipient PS, which then disappears at zero temperature
(besides our findings, for $T=0$ other different treatments find that the ground state is homogeneous
\cite{Hua03, Koc04, Cap04}). The nature of this reentrant behavior is still unclear. 

\begin{figure}[b]
\centering         %2DGammaq.eps
\includegraphics[width=8cm]{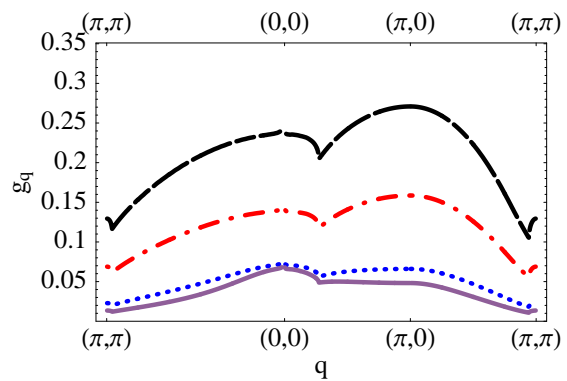}
\caption{2d-Renormalized $e$-$ph$ coupling for $n=0.9$ along the closed path $\qvec=(\pi,\pi) - (0,0) - (\pi,0)$ for $U/U_c =  0.46$  (dashed line), $U/U_c =  0.62$  (dotted-dashed line),  $U/U_c =  0.93$  (dotted line) and $U/U_c = 1.23$  (solid line). 
Results are given for a $1000 \times 1000$ lattice.}
\label{2DGammaq}
\end{figure}

In Fig.~\ref{2DGammaq} we show the vertex $g_\qvec$
%$\Gamma_\qvec = \beta \kappa_\qvec / \kappa^0_\qvec$ 
for different $U$ along a triangular path. 
This quantity displays minima at wavevectors $\qvec=\qvec_{c}$ (and at
$\qvec'_c$, see Fig. ~\ref{fig:2Dkq}) thus suppressing the
noninteracting instabilities. 
These minima arise because the bare susceptibility $\kappa^0_\qvec$ is maximal at these wavevectors, while
the corresponding quantity in the presence of $U$, $\kappa_\qvec$ is small 
due to the suppressed 
scattering at large momenta when the interaction becomes sizable. On the other hand the 
charge susceptibility is reduced less at small momenta and this gives rise to the pronounced 
maximum around $\qvec$=(0,0) in the large-$U$ case ($U\gtrsim 0.9$ in Fig.~\ref{2DGammaq}).
 The shape of the curves given in Fig.~\ref{2DGammaq} is  very similar
to those obtained within a SB calculations at $T=0.002$.\cite{Koc04} 
However, this seeming agreement has to be taken with a pinch of salt since
the results shown in Ref. \onlinecite{Koc04} are for the quasiparticle-phonon
vertex while ours correspond to the vertex for bare electrons and thus
should differ by a factor $z_0^2$. 
 
\section{Conclusions}
\label{sec:con}
In this work we have investigated the effects of strong electronic correlations on the $e$-$ph$ 
coupling, in particular we considered the case of phonons coupled to the local charge density as 
described by the Hubbard-Holstein model. We first exploited the adiabatic limit of the lattice degrees of freedom
to derive an exact result relating the screening of the $e$-$ph$ coupling to the purely 
electronic static charge susceptibility. This result holds generically for any kind of 
$e$-$e$ interaction (not only for the Hubbard one) and should also provide 
valuable  information in 
the partially adiabatic case of finite phonon frequency ($\omega_0\ll t$). 
It is important to
note that the analysis of  the correlation-driven screening of the  $e$-$ph$ 
coupling can be performed by investigating the purely electronic problem. 
The latter was investigated within  the static limit of the GA+RPA method.
This technique assumes a Fermi-liquid ground state and considers the low-energy quasiparticle 
physics. Therefore our low-energy description of the electron liquid is appropriate in high dimensions,
where the Fermi-liquid is a good starting point. Particularly favorable is the $d=\infty$ case,
where the GA becomes the  exact solution of the GZW variational problem.

 The main outcome is that (strong) 
correlations induce rigidity in the charge density
fluctuations thereby reducing the effective $e$-$ph$ coupling when this is of the Holstein type.
More specifically the analysis of the momentum dependence shows that the $e$-$ph$ coupling is more
severely reduced in processes with large momentum transfer. This result, which was already known
in large-N approaches to the infinite-$U$ Hubbard-Holstein models, 
\cite{Gri94, Kel95, Zey96} is 
considered here within a systematic variation of the correlation strength. 
In particular, from
Figs.~\ref{2Dgq} and \ref{2DGammaq}  one can see that, while at small $U$'s  the effective
$e$-$ph$ vertices at small and large transferred momenta differ at most
by 40 percent, at large $U$'s the $e$-$ph$ coupling at large momenta can be five or more
times smaller than the couplings at low momenta.
The fact that the $e$-$ph$ is screened less for small transferred
momenta has important consequences as far as the charge instabilities of the model 
are concerned. Indeed, the $e$-$e$ interaction not only generically reduces the effect of the 
phonons and enhances the minimum strength of $e$-$ph$ coupling to drive the system unstable,
but also introduces a momentum dependence, which changes the nature of the instability upon increasing
the strength of the correlation: While for small $e$-$e$ interaction the leading 
instability is of the Peierls type, with the formation
of CDW at momenta $|\qvec_c|=2k_F$, upon increasing $U$, the scattering processes at small transferred 
momentum become comparatively stronger and lead to a PS instability at vanishing $\qvec_c$.
Our technique allows a systematic investigation of how the low-coupling 
CDW instability transforms into the PS instability 
leading to a phase diagram like tha one shown in
Fig. \ref{2DPhDiaTot} for the 2-d system.

 Of course the PS  instability is specific to the short range nature
 of the model. When the long range Coulomb interaction is included the large-scale PS of charged holes
is prevented and a frustrated PS occurs with the formation of various possible textures 
\cite{CDCG95,Sei00,lor01I,lor02,Ort06,Ort07,ort08}.

We also notice that the bare $e$-$ph$ $\lambda$ needed to drive the systems unstable are rather small
(of order one or less) at large $U$ and moderate doping. A good compromise is indeed reached 
in this region, where the quasiparticles have a substantially reduced kinetic energy 
(the effective mass is 3-5 times larger than the bare one), but the system is not too close to
the insulating phase, where the interaction would screen too severely the $e$-$ph$ coupling.
Therefore, in this rather metallic regime the (frustrated) PS instability is quite competitive
with respect to the polaron formation, which could instead be favored by the stronger correlation 
effects occurring in the antiferromagnetic region of the phase diagram\cite{MCgunnarson}.

The general interest of the above findings and the encouraging reliability test of the GA technique
discussed in the present work are a stimulating support for the extension of the present work
to the dynamical regime. In this case future natural extensions will consider the analysis
of correlation effects on the phonon dynamics and their investigation in broken-symmetry states,
like the stripe phase. 

\acknowledgments
%We thank for illuminating discussions.
This work has been supported by MIUR PRIN05 (prot. 2005022492) and by CNR-INFM.
We also thank the VIGONI foundation for financial support.
\section*{APPENDIX}
%\label{sec:for}
\subsection*{Real space energy expansion}
Here we give a derivation of the second order term $\delta E_{e}^{(2)}$
in the energy expansion Eq. (\ref{exp}).
In real space we obtain

\begin{widetext}
\begin{eqnarray}
\delta E_{e}^{(2)} = \sum_{ij}t_{ij}&\{& \frac{1}{2}z_0 [z^{'}(\delta \rho_{ii} 
\delta T_{ij}^C + \delta m_i \delta T_{ij}^S) + z^{'}_{+-}(\delta \rho_{ii} \delta 
T_{ij}^C - \delta m_i \delta T_{ij}^S)] \nonumber \\ &+&  \frac{1}{8}T_{ij0} [(z^{'}+ 
z^{'}_{+-})^2 \delta \rho_{ii} \delta \rho_{jj} + (z^{'}- z^{'}_{+-})^2 \delta m_i 
\delta m_j +z_0 (z^{''}_{++} + 2z^{''}_{+-} + z^{''}_{--}) (\delta \rho_{ii})^2 \nonumber \\ 
&+&z_0(z^{''}_{++} - 2z^{''}_{+-} + z^{''}_{--}) (\delta m_i)^2]+\frac{1}{2}T_{ij0} 
[z^{'}_D (z^{'} + z^{'}_{+-})\delta \rho_{ii} \delta D_j + z_0 (z^{''}_{+D}+ z^{''}_{-D})
\delta \rho_{ii} \delta D_i ] \nonumber \\ &+& z_0 z^{'}_D \delta T_{ij}^C \delta D_i + 
\frac{1}{2}T_{ij0} [(z^{'}_D)^2 \delta D_i \delta D_j + z_0 z^{''}_D (\delta D_i)^2] \ \  \}
\label{Er4}
\end{eqnarray}
\end{widetext}

where the fluctuating variables correspond to
the charge density $\rho_{ii} = \rho_{ii\uparrow} + \rho_{ii\downarrow}$ 
and the magnetization density $m_{i} = \rho_{ii\uparrow} - \rho_{ii\downarrow}$.
Further on we have defined the transitive fluctuations
\begin{eqnarray*}
T_{ij} = T_{ij0} + \delta T_{ij}^{C} + \delta T_{ij}^{S}
\label{T}
\end{eqnarray*}
which in the charge- and spin sector read as 
\begin{eqnarray*}
&&  \delta T_{ij}^{C} = \sum_{\sigma} (\delta\langle c^{\dag}_{i \sigma}c_{j \sigma}\rangle  + 
\delta\langle c^{\dag}_{j \sigma}c_{i \sigma}\rangle ) \nonumber  \\ && \delta T_{ij}^{S} = 
\sum_{\sigma} \sigma (\delta\langle c^{\dag}_{i \sigma}c_{j \sigma}\rangle  + \delta\langle 
c^{\dag}_{j \sigma}c_{i \sigma}\rangle ).
\end{eqnarray*}

Since we study a paramagnetic system it is convenient to define the
following abbreviations for the z-factors and its derivatives:

\begin{eqnarray*}
 && z_{i\sigma}\equiv z_0 , \ \ \frac{\partial z_{i \sigma}}{\partial \rho_{ii \sigma}}\equiv z^{'},  \nonumber \\ &&  \frac{\partial z_{i\sigma}}{\partial \rho_{ii -\sigma}}\equiv z^{'}_{+-}, \ 
\frac{\partial z_{i \sigma}}{\partial D_{i}}\equiv z^{'}_{D}
\label{z1} \\
&& \frac{\partial^2 z_{i \sigma}}{\partial \rho^{2}_{ii \sigma}} \equiv z^{''}_{++}, \  
\frac{\partial^2 z_{i \sigma}}{\partial \rho_{ii \sigma}\partial \rho_{ii -\sigma}}
\equiv z^{''}_{+-}, \ \frac{\partial^2 z_{i \sigma}}{\partial \rho^{2}_{ii -\sigma}} \equiv z^{''}_{--}  \nonumber \\ 
&&  \frac{\partial^2 z_{i \sigma}}{\partial D^{2}_{i}}\equiv z^{''}_{D}, \ \frac{\partial^2 
z_{i \sigma}}{\partial \rho_{ii \sigma}\partial D_{i}}\equiv z^{''}_{+D}, \ \frac{\partial^2 
z_{i \sigma}}{\partial \rho_{ii -\sigma}\partial D_{i}}\equiv z^{''}_{-D}
\label{z2}
\end{eqnarray*} 
For the half-filled paramagnetic state we have  $z^{'} = z^{'}_{+-}$
and $z^{''}_{+D}=z^{''}_{-D}$.

\subsection*{Momentum space energy expansion}
We transform Eq.~(\ref{Er4}) into momentum space. For the paramagnetic
system the expansion separates into the charge- and spin sector 
$\delta E_{e}^{(2)} = \delta E_{e}^{S} + \delta E_{e}^{C}$.

In the spin sector we find: 
\begin{eqnarray}
\delta E_{e}^S = \frac{1}{N} \sum_{\qvec}&[&\frac{1}{2}z_0 (z^{'} - z^{'}_{+-})(\delta 
S^{z}_\qvec \delta T^{S}_{-\qvec} + \nonumber \\ &+& \delta T^{S}_\qvec \delta S^{z}_{-\qvec}) + 
N_\qvec \delta S^{z}_\qvec \delta S^{z}_{-\qvec} \ \ ] \label{ES}
\end{eqnarray}
with the following definitions:
\begin{eqnarray*}
\delta m_{i}=\frac{1}{N}\sum_{\qvec}e^{i\qvec_i\cdot \rvec_i}\delta m_{\qvec}=\frac{2}{N}
\sum_{\qvec}e^{i\qvec_i\cdot\rvec_i} \delta S^{z}_{ \qvec}
\label{dm}
\end{eqnarray*}
\begin{eqnarray*}
N_\qvec = \frac{1}{N} &[&(z^{'}-z^{'}_{+-})^{2} \sum_{\kvec\sigma}\epsilon^{0}_{\kvec+\qvec,\sigma}
n_{\kvec\sigma} \nonumber \\ &+& z_{0}(z^{''}_{++} - 2z^{''}_{+-}+z^{''}_{--})\sum_{\kvec\sigma}
\epsilon^{0}_{\kvec\sigma}n_{\kvec\sigma} \ \ ]
\label{Nq}
\end{eqnarray*}
The form of $\delta E_{e}^C$ in momentum space is slightly more complicate:
\begin{widetext}
\begin{eqnarray}
\delta E_{e}^C &=& \frac{1}{2N}z_{0}(z^{'}+z^{'}_{+-})\sum_{\qvec}\delta \rho_{\qvec} \delta T^{C}_{-\qvec} \nonumber \\  
&+&  \frac{1}{4N^{2}}\sum_{\kvec \qvec\sigma}[(z^{'}+z^{'}_{+-})^{2} \epsilon^{0}_{\kvec+\qvec,\sigma}n_{\kvec\sigma} 
+ (z^{''}_{++} + 2z^{''}_{+-}+z^{''}_{--}) z_{0}\epsilon^{0}_{\kvec\sigma}n_{\kvec\sigma}]\delta 
\rho_{\qvec}\delta \rho_{-\qvec} \nonumber \\  &+&  \frac{1}{N^{2}}\sum_{\kvec \qvec\sigma}[z^{'}_{D}(z^{'}+z^{'}_{+-}) 
\epsilon^{0}_{\kvec+\qvec,\sigma}n_{\kvec\sigma} + z_{0}(z^{''}_{+D} + z^{''}_{-D})]\delta \rho_{\qvec} \delta D_{-\qvec} 
\nonumber \\  &+&  \frac{1}{N}\sum_{\qvec}z_{0}z^{'}_D \delta D_{\qvec} \delta T^{C}_{-\qvec} + 
\frac{1}{N^{2}}\sum_{\kvec \qvec\sigma}[(z^{'}_{D})^{2}\epsilon^{0}_{\kvec+\qvec,\sigma}n_{\kvec\sigma} + 
z_{0}z^{''}_{D} \epsilon^{0}_{\kvec\sigma}n_{\kvec\sigma}]\delta D_{\qvec} \delta D_{-\qvec}
\label{Ek01}
\end{eqnarray}
\end{widetext}
From Eq.~(\ref{Ek01}), we recover Eq.~(\ref{eskspace}) by 
using the definitions of $L_\qvec$, $U_\qvec$ and $V_\qvec$ from Eq. \ref{eq:vlu}.

%\bibliographystyle{prsty_no_etal}
%\bibliography{htsct,htsce,htsct-marco,htsce-marco}
%%%%% THIS IS THE BIBLIOGRAPHY GENERATED WITH BIBTEX AND THE ABOVE
%%%%% COMMANDS.

 \end{document}